\documentclass[prl,aps,showpacs,twocolumn,superscriptaddress,notitlepage]{revtex4}
\usepackage{epsfig,amsfonts}
\usepackage{amsmath}
\usepackage{bbm}
\usepackage{graphicx,psfrag,rotating}
\usepackage{graphics}
\usepackage{txfonts}

\usepackage{graphicx}
\usepackage{dcolumn}
\usepackage{bm}

\begin{document}

\title{Can coarse measurements reveal macroscopic quantum effects?}

\author{Tian Wang}
\address{Institute for Quantum Science and Technology, Department of Physics and Astronomy, University of Calgary, Alberta T2N 1N4, Canada}

\author{Roohollah Ghobadi}
\address{Institute for Quantum Science and Technology,
Department of Physics and Astronomy, University of Calgary, Alberta T2N 1N4, Canada}

\author{Sadegh Raeisi}
\address{Institute for Quantum Computing, University of Waterloo, Ontario N2L 3G1, Canada}
\author{Christoph Simon}
\address{Institute for Quantum Science and Technology,
Department of Physics and Astronomy, University of Calgary, Alberta T2N 1N4, Canada}

\begin{abstract}
It has recently been conjectured that detecting quantum effects such as superposition or entanglement for macroscopic systems always requires high measurement precision. Analyzing an apparent counter-example involving macroscopic coherent states and Kerr non-linearities, we find that while measurements with coarse outcomes can be sufficient, the phase control precision of the necessary non-linear operations has to increase with the size of the system. This suggests a refined conjecture that either the {\it outcome precision} or the {\it control precision} of the measurements has to increase with system size.
\end{abstract}
\maketitle

What does it take to observe quantum effects such as superposition and entanglement for macroscopic systems? It is essential to isolate the system well from its environment in order to suppress decoherence \cite{zurek}. However, there are several results that suggest that this is not sufficient, and that the precision of the measurements that one is able to perform on the system also plays an important role. Mermin \cite{mermin} showed in 1980 that in order to obtain a Bell inequality violation for singlet states of two large spins $s$, the directions of the spin measurements had to be chosen with an angular resolution that increased with the size of the spins as $1/s$. Note that here and in the following we speak of 'increasing' resolution or precision when the acceptable error or uncertainty decreases. The requirement of choosing the direction precisely is an example for necessary measurement {\it control precision}, i.e. the precision with which relevant physical parameters have to be controlled in order to implement the desired measurement procedure.

Later Peres \cite{peres} showed that for the same singlet state of two spins the precision with which the measurement outcomes are known is also important. He showed that if this measurement {\it outcome precision} is worse than $O(\frac{1}{\sqrt{s}})$ in relative terms, then a classical model can reproduce the quantum predictions for the correlation functions. Related results for individual large spins were obtained in Ref. \cite{kofler}. Ref. \cite{entlaser} studied multi-photon singlet states equivalent to Mermin's and Peres' spin singlets and showed that $O(\frac{1}{\sqrt{N}})$ relative outcome precision (where $N$ is the photon number) is sufficient to demonstrate entanglement. Most recently Ref. \cite{raeisi1} studied so-called micro-macro entangled states of light that are obtained by greatly amplifying one half of an initial entangled photon pair. These authors found that a relative outcome precision of order $\frac{1}{N}$ was necessary to see quantum effects in this example. Similar results on the effect of coarse-graining on macroscopic entanglement were found in Refs. \cite{spagnolo,portolan}.

Ref. \cite{raeisi1} also put forward the conjecture that demonstrating quantum effects in macroscopic systems always requires high measurement precision.
In contrast, Ref. \cite{jeongcoarse} proposed a state and measurement procedure based on the use of Kerr non-linearities where a Bell inequality violation could apparently be observed with very coarse measurements. As a first step towards addressing this apparent contradiction, Ref. \cite{raeisi1} pointed out that the non-linear operations used in the proposal of Ref. \cite{jeongcoarse} involve large ($\pi$) phase shifts between neighboring Fock states and suggested that this could be seen as high resolution in a more general sense. Later Ref. \cite{ghobadi} showed that in order to prepare entangled states of the type used in Ref. \cite{jeongcoarse} the phase of the non-linear operations has to be controlled with a precision that increases with system size. Ref. \cite{ghobadi} is linked to the present work in that it already highlighted the importance of phase precision. However, it focused on state preparation. Here we explicitly address the question of measurement precision posed in Ref. \cite{raeisi1}. We show that even if one assumes that the states under consideration are ideal, measurement precision - in particular control precision - has to increase with system size in order to be able to demonstrate quantum effects.

We study superpositions and entanglement involving coherent states with opposite amplitudes, $|\alpha\rangle$ and $|-\alpha\rangle$, where we will take $\alpha$ to be real for simplicity. We will pay particular attention to the macroscopic limit $\alpha \gg 1$. We study this example not only because these states lie at the heart of the proposal of Ref. \cite{jeongcoarse}, but also because they are a well-known ``archetype'' for macroscopic quantum superpositions \cite{yurke,cats,schoelkopf}. Let us note right away that the proposal of Ref. \cite{jeongcoarse} is more complex than the simple cases considered here. However, our conclusions concerning control precision apply to that work as well. We focus on simple states and measurement schemes for clarity.

\begin{figure}
\includegraphics[width=\columnwidth]{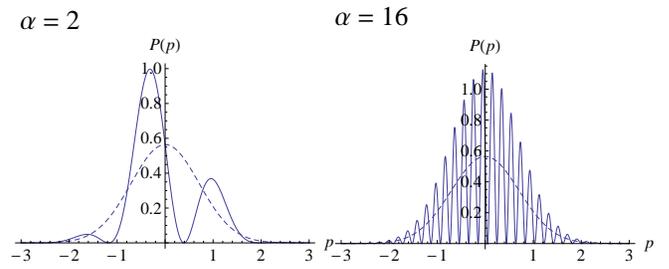}
\caption{Probability of outcomes for measurements of the $\hat{p}$ quadrature for the superposition state of Eq. (\ref{alpha+}) and the mixed state of Eq. (\ref{rho}) for $\alpha=2$ (left) and $\alpha=16$ (right). The oscillatory structure that distinguishes the two distributions becomes harder to resolve as $\alpha$ increases, see also Eqs. (\ref{palpha+}) and (\ref{prho}).}
\label{oscillation}
\end{figure}

We begin by considering the superposition state
\begin{equation}
|\alpha_+\rangle=\frac{1}{\sqrt{2}} (|\alpha\rangle+i|-\alpha\rangle),
\label{alpha+}
\end{equation}
 focusing on the regime where $\alpha$ is large enough such that the overlap $\langle \alpha|-\alpha\rangle=e^{-2 \alpha^2}$ is negligible. The phase factor $i$ is chosen for convenience. This state can be created, for example \cite{yurke,schoelkopf}, from an initial coherent state with the help of a Kerr nonlinearity,
\begin{equation}
e^{-i\frac{\pi}{2} \hat{N}^2}|\alpha\rangle = e^{-i\frac{\pi}{4}}|\alpha_+\rangle,
\label{makealphaplus}
\end{equation}
where $\hat{N}=a^{\dagger}a$, and $a$ is the bosonic annihilation operator for which the coherent state is an eigenstate, $a|\alpha\rangle=\alpha|\alpha\rangle$. It was shown in Ref. \cite{ghobadi} that the phase of the unitary operation in Eq. (\ref{makealphaplus}) has to be precisely equal to $\frac{\pi}{2}$ in order to generate this state with high fidelity, with a precision that increases with $\alpha$. However, as mentioned in the introduction, this is not our concern here. We will assume that the ideal state is given to us and focus on the question of how to prove that we have a quantum superposition state, as opposed to a ``classical'' mixture of the same two coherent states,
\begin{equation}
\rho=\frac{1}{2}(|\alpha\rangle \langle \alpha|+|-\alpha\rangle \langle -\alpha|).
\label{rho}
\end{equation}
Let us first consider measurements of the quadrature $\hat{x}=\frac{1}{2}(a+a^{\dagger})$. For the state of Eq. (\ref{alpha+}), this will give a symmetric bimodal distribution of results corresponding to the two components of the superposition,
\begin{equation}
P(x)=|\langle x|\alpha_+\rangle|^2=\frac{e^{-(x+\alpha)^2}+e^{-(x-\alpha)^2}}{2\sqrt{\pi}},
\label{Px}
\end{equation}
where $\hat{x}|x\rangle=x|x\rangle$. Note that for $\alpha\gg 1$ one can distinguish the two components using very coarse measurements of $\hat{x}$; this point will be significant below. However, this does not prove that one is dealing with a macroscopic superposition state, since the mixed state of Eq. (\ref{rho}) will produce the exact same distribution of outcomes. In general, one has to measure at least two {\it non-commuting} observables in order to prove the quantum character of any system. One obvious choice for an observable that does not commute with $\hat{x}$ is the complementary quadrature, $\hat{p}=\frac{-i}{2}(a-a^{\dagger})$. The probability distribution of the associated outcomes $p$ is
\begin{equation}
P_{|\alpha+\rangle}(p)=|\langle p|\alpha_+\rangle|^2=\frac{e^{-p^2}(1-\sin(2\alpha p))}{\sqrt{\pi}}
\label{palpha+}
\end{equation}
where  $\hat{p}|p\rangle=p|p\rangle$, whereas for the mixed state of Eq. (\ref{rho}) one has
\begin{equation}
P_{\rho}(p)=\langle p|\rho|p\rangle=\frac{e^{-p^2}}{\sqrt{\pi}}
\label{prho}
\end{equation}
 The two probability distributions are different, which means that the measurement of $\hat{p}$ can indeed be used to discriminate Eq. (\ref{alpha+}) from Eq. (\ref{rho}). However, the difference is due to the oscillatory term in Eq. (\ref{palpha+}), whose oscillation frequency increases with increasing $\alpha$. Detecting this oscillation therefore requires a precision in the $\hat{p}$ measurement that increases with $\alpha$, see also Fig. \ref{oscillation}. In fact, this was one of the examples mentioned in Ref. \cite{raeisi1} in order to argue for the plausibility of the considered conjecture. The same effect can also be discussed in terms of the Wigner function \cite{zureknature}. Fig. \ref{oscillation} could also be compared to Fig. 2 of Ref. \cite{raeisi1}, which shows a similar effect for a different macroscopic quantum state.

\begin{figure}
\includegraphics[width=\columnwidth]{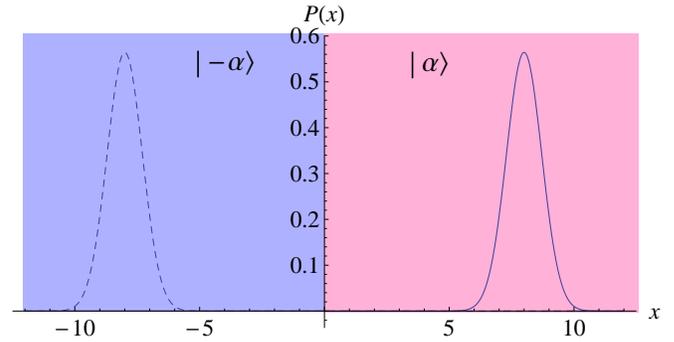}
\caption{Outcome distributions for measurements of the $\hat{x}$ quadrature for the states $|\alpha\rangle$ (solid) and $|-\alpha\rangle$ (dashed) for $\alpha=8$. For large enough $\alpha$, the two states can be distinguished by a very coarse measurement. Positive values (red) of $\hat{x}$ can be assigned to $|\alpha\rangle$ and negative values (blue) to $|-\alpha\rangle$. The overlap between the two distributions, and thus the error of this measurement scheme, is negligible.}
\label{coarse}
\end{figure}

However, there is a different approach to proving the superposition character of Eq. (\ref{alpha+}), which is closely linked to the proposal of Ref. \cite{jeongcoarse}. One can view the states $|\alpha\rangle$ and $|-\alpha\rangle$ as the computational basis states of a ``coherent state qubit'' \cite{alphaqbits,jeongkim}. Measurements in the computational basis, which we will also refer to as $\sigma_z$ measurements (where $\sigma_z=|\alpha\rangle \langle \alpha|-|-\alpha\rangle \langle -\alpha|$), can clearly be done in a very coarse way, e.g. by measuring $\hat{x}$. For large enough $\alpha$, positive (negative) values correspond to the state $|\alpha\rangle$ ($|-\alpha\rangle$) with extremely high fidelity, and coarse-graining the $x$ values only has a negligible effect on the measurement fidelity, see also Fig. \ref{coarse}.

As before, proving the quantum character of (\ref{alpha+}) requires at least one other measurement that does not commute with $\sigma_z$. A natural choice from the qubit perspective is
\begin{equation}
\sigma_y=|\alpha_+\rangle \langle \alpha+|-|\alpha_-\rangle \langle \alpha_-|,
\end{equation}
where $|\alpha_-\rangle=\frac{1}{\sqrt{2}}(i|\alpha\rangle+|-\alpha\rangle$. If $\sigma_y$ can be measured, then it is obviously easy to prove that a given source produces the state Eq. (\ref{alpha+}) - the corresponding measurement will always give the result $+1$ and never $-1$, whereas for the mixed state (\ref{rho}) the results would be 50/50.

\begin{figure}
\includegraphics[width=\columnwidth]{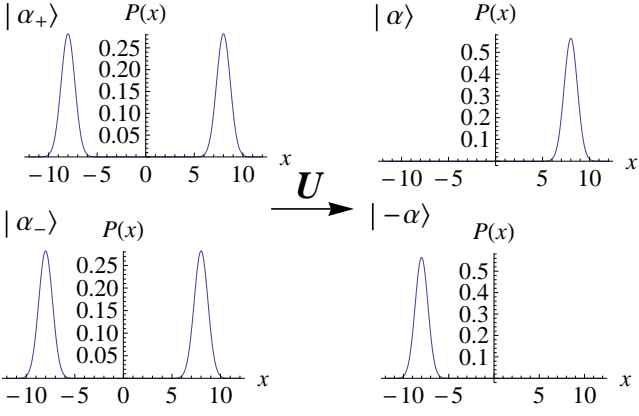}
\caption{The $\hat{x}$ quadrature distributions for the states $|\alpha_+\rangle=\frac{1}{\sqrt{2}}(|\alpha\rangle+i|-\alpha\rangle$ (top left) and $|\alpha_-\rangle=\frac{1}{\sqrt{2}}(i|\alpha\rangle+|-\alpha\rangle$ (bottom left) are identical. However, application of the Kerr rotation Eq. (\ref{U}) transforms $|\alpha_+\rangle$ into $|\alpha\rangle$ (top right) and $|\alpha_-\rangle$ into $|-\alpha\rangle$ (bottom right). These states can now be distinguished by a coarse measurement as in Fig. \ref{coarse}.}
\label{Ufig}
\end{figure}

The required measurement of $\sigma_y$ can be implemented using a Kerr non-linearity, see also Ref. \cite{jeongkim}. Changing the sign of $\alpha$ in Eq. (\ref{makealphaplus}) one has $e^{-i\frac{\pi}{2} \hat{N}^2}|-\alpha\rangle = e^{-i\frac{\pi}{4}}|\alpha_-\rangle$. Inverting these relations one sees that the Kerr operation allows one to rotate the $\sigma_y$ eigenstates into the $\sigma_z$ eigenstates, i.e.
\begin{equation}
U|\alpha_+\rangle=|\alpha\rangle, U|\alpha_-\rangle=|-\alpha\rangle,
\end{equation}
where
\begin{equation}
U=e^{-i\frac{\pi}{4}} e^{i \frac{\pi}{2} \hat{N}^2}.
\label{U}
\end{equation}
This means that a measurement of $\sigma_y$ can be done on an arbitrary state by first applying the rotation $U$, followed by a measurement in the $\sigma_z$ basis, as shown in Fig. \ref{Ufig}. As mentioned before and in Fig. \ref{coarse}, the $\sigma_z$ measurement can be done in a very coarse way. This means that it is possible to prove the presence of the macroscopic superposition (\ref{alpha+}) using measurements that are coarse in terms of outcome resolution.

\begin{figure}
\includegraphics[width=\columnwidth]{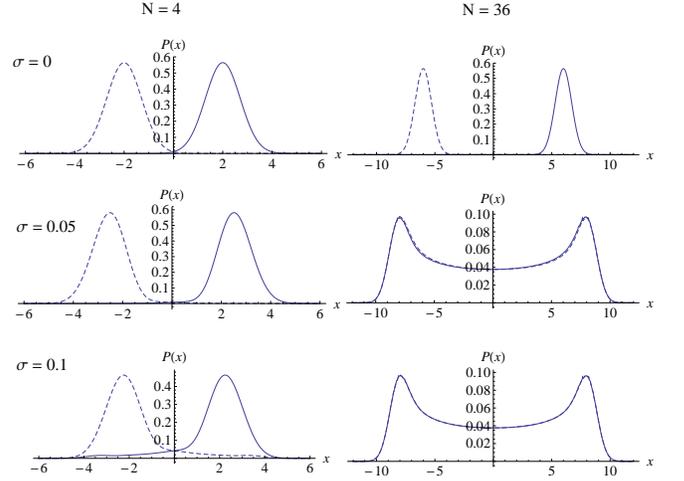}
\caption{Outcome distributions for $\hat{x}$ quadrature measurements for the states $C_{\sigma}(|\alpha\rangle \langle \alpha|)$ (solid) and $C_{\sigma}(|-\alpha\rangle \langle -\alpha|)$ (dashed) that are created from the states $|\alpha_+\rangle$ and $|\alpha_-\rangle$ by a Kerr rotation with Gaussian phase uncertainty $\sigma$, see Eq. (\ref{average}). We show the case $N=\alpha^2=4$ on the left and $N=36$ on the right, with $\sigma$ increasing from top to bottom. One sees that the distributions overlap much faster for greater $N$, leading to errors in the $\sigma_y$ measurement of Fig. \ref{Ufig}, see also Fig. \ref{coarse}. For large enough $\sigma$ it becomes impossible to distinguish the two states.}
\label{sigmaz}
\end{figure}

However, we argue that it is physically important to also consider the necessary control precision. The control parameter that we focus on here is the phase of the Kerr rotation $U$. Suppose that instead of exactly $\frac{\pi}{2}$ this phase is $\frac{\pi}{2}+\phi$. Then, when trying to perform the $\sigma_y$ measurement, the state $|\alpha_+\rangle$ will be rotated not into $|\alpha\rangle$, but into $e^{i\phi\hat{N}^2}|\alpha\rangle$, and $|\alpha_-\rangle$ into $e^{i\phi\hat{N}^2}|-\alpha\rangle$. For simplicity let us consider a Gaussian distribution for $\phi$ with a width $\sigma \ll 1$ (which is the relevant regime, as will become clear below). Then the final state corresponding to $|\alpha_+\rangle$ is
\begin{eqnarray}
C_{\sigma}(|\alpha\rangle \langle \alpha|)=\frac{1}{\sqrt{2\pi}\sigma}\int_{-\infty}^{\infty} d\phi e^{-\frac{1}{2}\frac{\phi^2}{\sigma^2}} e^{i\phi\hat{N}^2}|\alpha\rangle \langle \alpha|e^{-i\phi \hat{N}^2}=\nonumber\\
\frac{e^{-\alpha^2}}{\sqrt{2\pi}\sigma}\int_{-\infty}^{\infty} d\phi e^{-\frac{1}{2}\frac{\phi^2}{\sigma^2}} \sum_{n,n'=0}^{\infty} e^{i\phi(n^2-n'^2)} \frac{\alpha^{n+n'}}{\sqrt{n!}\sqrt{n'!}} |n\rangle \langle n'|,
\label{average}
\end{eqnarray}
where we have introduced the notation $C_{\sigma}$ for the associated error channel, extended the range of integration for $\phi$ to infinity (which can be done with negligible error for $\sigma \ll 1$), and expanded $|\alpha\rangle$ in terms of photon number states. Performing the integration over $\phi$ one finds
\begin{equation}
e^{-\alpha^2}\sum_{n,n'=0}^{\infty} e^{-\frac{1}{2} \sigma^2 (n^2-n'^2)^2} \frac{\alpha^{n+n'}}{\sqrt{n!}\sqrt{n'!}} |n\rangle \langle n'|.
\label{averagedalpha}
\end{equation}
The term containing $\sigma$ leads to a suppression of the off-diagonal elements in the number state basis. The key point for the present work is that this suppression happens faster for larger values of $\alpha$. This can be seen by remembering that the number distribution for a coherent state is a Poissonian with a peak at $\alpha^2$ (and a corresponding width $\alpha$). For large enough $\alpha$ one can then approximate the factor $(n^2-n'^2)^2=(n+n')^2 (n-n')^2$ in the exponential in Eq. (\ref{averagedalpha}) by $4 \alpha^4 (n-n')^2$. This shows that the off-diagonal elements are suppressed by a Gaussian factor $e^{-2 \sigma^2 \alpha^4 (n-n')^2}$. This means that for $\sigma \alpha^2 \gtrsim 1$ the state (\ref{averagedalpha}) is essentially diagonal in the number basis. Moreover the state corresponding to $|\alpha_-\rangle$, which we denote $C_{\sigma}(|-\alpha\rangle \langle -\alpha|)$, converges to the same diagonal form. In this regime there is therefore no way to distinguish these two states, see also Fig. \ref{sigmaz}.

This means that the described procedure for measuring $\sigma_y$ breaks down for phase errors $\sigma$ that are of order $\frac{1}{\alpha^2}$, or $\frac{1}{N}$, if $N=\alpha^2$ is used to denote the typical number of particles in the system. The precision with which $\phi$ has to be controlled thus increases with system size.
The coherent state qubit approach relies on being able to confine the dynamics of the system to the two-dimensional subspace spanned by $|\alpha\rangle$ and $|-\alpha\rangle$, even though the number of Fock states that effectively contribute to the dynamics is of order $\alpha$ (due to the Poisson distribution of numbers for coherent states). This becomes more and more difficult for increasing $\alpha$. The evolution of coherent states under small Kerr rotations is discussed also in different terms in Refs. \cite{yamamoto,haroche}.

\begin{figure}
\includegraphics[width=\columnwidth]{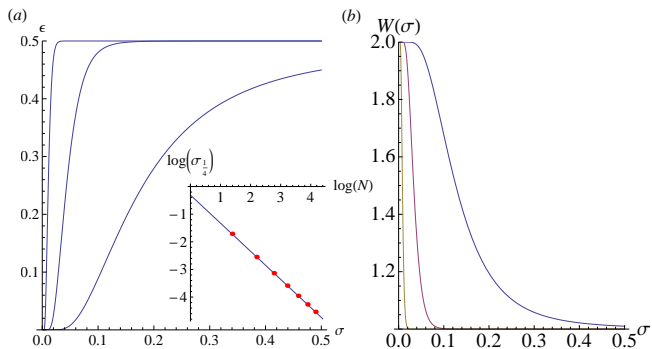}
\caption{(a) The bit-flip error $\epsilon$ in the $\sigma_y$ measurement of Fig. \ref{Ufig} as a function of the Kerr phase uncertainty $\sigma$, for the cases $N=\alpha^2=4, 16, 64$ from bottom to top. One sees that $\epsilon$ approaches $\frac{1}{2}$ for increasing $\sigma$, and this happens faster for greater $N$. The log-log plot in the inset shows that the value of $\sigma$ for which $\epsilon=\frac{1}{4}$ (i.e. half its asymptotic value) scales like $\frac{1}{N}$, as expected from the analytical argument given in the text. (b) Expectation value of the entanglement witness $W$ of Eq. (\ref{W}) for the state of Eq. (\ref{Phi-}), for $N=\alpha^2=4, 16, 64$ from top to bottom. For increasing $\sigma$ the value of $W$ approaches 1 (the bound for separable states), due to the bit-flip errors in the $\sigma_y$ measurement shown in (a). This happens faster for greater values of $N$. }
\label{epsilon}
\end{figure}

 This result holds no matter how the final measurement in the $\sigma_z$ basis is performed. For concreteness, we show in Figs. \ref{sigmaz} and \ref{epsilon}(a) how the phase error $\sigma$ affects the measurement strategy described in Figs. \ref{Ufig} and \ref{coarse}. Fig. \ref{sigmaz} shows that the $\hat{x}$ quadrature distributions for the states $C_{\sigma}(|\alpha\rangle \langle \alpha|)$ and $C_{\sigma}(|-\alpha\rangle \langle -\alpha|)$ begin to overlap for increasing $\sigma$, and that this happens much faster for greater values of $\alpha$. Fig. \ref{epsilon}(a) shows the resulting bit-flip error $\epsilon$ for the $\sigma_y$ measurement of Fig. \ref{Ufig}. This can be calculated as $\epsilon=\int_{-\infty}^0 dx \, P(x)$, with $P(x)$ the $\hat{x}$ quadrature distribution for the state $C_{\sigma}(|\alpha\rangle \langle \alpha|)$. As expected from the above discussion, $\epsilon$ approaches $\frac{1}{2}$ (corresponding to complete indistinguishability of the two states) for increasing $\sigma$, and this happens faster for greater values of $\alpha$.

So far we have discussed macroscopic superposition states. We now turn to the detection of macroscopic entanglement. Consider the state
\begin{equation}
|\Phi_-\rangle=\frac{1}{\sqrt{2}}(|\alpha\rangle |\alpha\rangle - |-\alpha\rangle |-\alpha\rangle),
\label{Phi-}
\end{equation}
where the relative sign between the two terms is chosen for convenience. This state can be created, for example, using a Kerr non-linearity, combined with a beam splitter and phase space displacements \cite{ghobadi,jeongkim,sanders}. Again our focus here is not on how to create the state, but on whether its entanglement can be demonstrated by coarse measurements.

As before, coarse quadrature measurements alone are not sufficient, but the coherent state qubit approach using the Kerr nonlinearity can be applied to the present case as well. Again measurements only in the computational basis ($\sigma_z$) are not sufficient to distinguish the entangled state (\ref{Phi-}) from a separable state, in particular from the 50/50 mixture of the product states $|\alpha\rangle |\alpha\rangle$ and $|-\alpha\rangle |-\alpha\rangle$. However, the entanglement can be demonstrated using the witness operator
\begin{equation}
W=\sigma_y \otimes \sigma_y + \sigma_z \otimes \sigma_z.
\label{W}
\end{equation}
One easily verifies that $\langle \Phi_-|W|\Phi_-\rangle=2$, whereas the modulus of the mean value of $W$ for separable states is bounded by one. This follows from the fact that for any state $|\chi\rangle$ the norm of the two-dimensional vector $\{\langle \chi |\sigma_y |\chi\rangle,\langle \chi| \sigma_z|\chi\rangle \}$ is bounded by one. For any product state, the mean value of $W$ is the scalar product of two such vectors, and its modulus is therefore also bounded by one; and every separable state is a convex combination of product states, thus satisfying the same bound, see also Ref. \cite{eisenberg}.

By performing measurements of $\sigma_z$ and $\sigma_y$ on each subsystem one can therefore prove the entanglement in the state (\ref{Phi-}). As discussed above, a coarse measurement of the $\hat{x}$ quadrature, for example, is sufficient to do the $\sigma_z$ measurement, but the $\sigma_y$ measurement requires moreover the Kerr rotation (\ref{U}). Therefore the exact same control precision requirements as above apply here as well. We showed in Fig. 5(a) that for a phase error $\sigma \gtrsim \frac{1}{\alpha^2}$, the bit-flip error $\epsilon$ in the $\sigma_y$ measurement  approaches $\frac{1}{2}$. The measured mean value of $W$, which is equal to $1+(1-2\epsilon)^2$ (as can easily be shown, assuming that the $\sigma_z$ measurement is perfect), therefore tends to 1, see also Fig. \ref{epsilon}(b). This means that the macroscopic entanglement becomes increasingly hard to detect as $\alpha$ increases. Note that as long as the mean value is greater than one, entanglement can in principle be proven. Our main point here is the scaling with $\alpha$. Due to this scaling, for any given non-zero level of experimental imperfection, there is a system size above which entanglement is no longer measurable.

We have seen that using macroscopic ``coherent state qubits'' one can in principle observe macroscopic quantum features such as superposition and entanglement using measurements that are very coarse in terms of outcome precision. However, there is a price to be paid. The measurements rely on being able to perform a rotation of the macroscopic qubit basis. When this rotation is implemented using a Kerr non-linearity, the control precision of the Kerr phase shift has to increase with the size of the system. The apparent counter-example of Ref. \cite{jeongcoarse} has thus led us to a refined formulation of the conjecture of Ref. \cite{raeisi1} that is both more precise and more general: the measurement precision required for demonstrating macroscopic quantum effects seems to increase with the size of the system, provided that both outcome precision and control precision are taken into account. This could be compared, for example, to the results of Ref. \cite{kofler2}, which studied the effect of coarse graining on macroscopic realism as defined by Leggett \cite{leggett} and emphasized the computational complexity (rather than the precision) of the operations that were required to observe violations of macroscopic realism.

The above conjecture is attractive, but it is far from proven. Different parts of our argument have a different degree of generality. The requirement for a rotation from a macroscopic superposition basis to a ``computational'' basis is very general in the present context. On the one hand, for a coarse measurement approach to work there has to be one basis for which the relevant states are easy to distinguish. On the other hand, to prove quantum characteristics one also has to be able to measure at least one observable that corresponds to a different basis, hence the need for a rotation between that basis and the computational basis. As a simple extension, one might want to consider other superposition states or entangled states using the same coherent-state qubits. Proving superpositions or entanglement  then requires slightly different rotations. More general qubit basis rotations can be constructed out of the Hadamard-type rotation $U$ of Eq. (\ref{U}) and phase space displacements \cite{jeongkim}. The same control precision requirements apply for this construction. They also apply to the measurements proposed in Ref. \cite{jeongcoarse}.

But could there be other ways of performing the basic Hadamard rotation? Do they necessarily have the same control precision requirements? In fact, it is known that the Kerr non-linearity is not the only possible solution \cite{yurke}. Higher powers of $\hat{N}^2$ also work. However, by adapting the argumentation around Eqs. (\ref{average},\ref{averagedalpha}) to these cases one can easily show that the control precision requirements are only increased in this case. For a Hamiltonian proportional to $\hat{N}^{2k}$ the necessary control precision scales as $\frac{1}{N^{2k-1}}$. So the Kerr non-linearity is optimal at least for this family of possible approaches.

We suggest that the basic difficulty with implementing a macroscopic basis rotation of the type of Eq. (\ref{U}) stems from the fact that the underlying Hilbert space is very large. In our case the effective Hilbert space dimension is of order $\alpha$, corresponding to the range of photon numbers that have significant weights for a coherent state. For increasing $\alpha$ it requires more and more fine-tuning to perform a non-trivial operation on the states $|\alpha\rangle$ and $|-\alpha\rangle$, while confining them to the two-dimensional subspace that they span. This may be a generic difficulty for macroscopic quantum systems.

We feel that proving these conjectures and intuitions would be very interesting, as it would significantly advance our understanding of the macroscopic limit of quantum physics. It would possibly be even more interesting if one could find a counter-example, since the latter might provide a promising avenue towards the demonstration of truly macroscopic quantum effects.

This work was supported by AITF and NSERC.



\end{document}